\documentclass[
twocolumn,
prb,
amsmath,
amssymb,
aps,
longbibliography,
superscriptaddress
]{revtex4-1}

\usepackage{graphicx,lineno,bm,mathtools,float}
\usepackage[colorlinks=true, allcolors=blue]{hyperref}
\begin{document}
	
	\title{Equilibrium magnetization of a quasispherical cluster of single-domain particles}
	
	\author{Andrey A. Kuznetsov}
	\email{kuznetsov.a@icmm.ru}
	\affiliation{
		Institute of Continuous Media Mechanics UB RAS,
		Perm Federal Research Center UB RAS,
		614013, Perm, Russia
	}

\begin{abstract}
Equilibrium magnetization curve of a rigid finite-size spherical cluster 
of single-domain particles is investigated both numerically and analytically. 
The spatial distribution of particles within the cluster is random. 
Dipole-dipole interactions between particles are taken into account. 
The particles are monodisperse. 
It is shown, using the stochastic Landau-Lifshitz-Gilbert equation, 
that the magnetization of such clusters is generally lower than predicted 
by the classical Langevin model. 
In a broad range of dipolar coupling parameters and particle volume fractions,
the cluster magnetization in the weak field limit can be successfully described 
by the modified mean-field theory, 
which was originally proposed for the description of concentrated ferrofluids.
In moderate and strong fields, the theory overestimates the cluster magnetization. 
However, predictions of the theory can be improved by adjusting the corresponding 
mean-field parameter. 
If magnetic anisotropy of particles is additionally taken into account 
and if the distribution of the particles' easy axes is random and uniform,
then the cluster equilibrium response is even weaker.
The decrease of the magnetization with increasing anisotropy constant is
more pronounced at large applied fields.
The phenomenological generalization of the modified mean-field theory,
that correctly describes this effect for small coupling parameters, 
is proposed.
\end{abstract}

\maketitle

\section{Introduction}

Nano- and micro-sized assemblies of single-domain 
particles are of great interest in modern biotechnology
and medicine. Prominent examples are composite magnetic microspheres 
(or ``magnetic beads'') which consist of fine magnetic particles
dispersed in or layered onto a spherical (usually polymer or silica) matrix~\cite{gijs2004magnetic,gervald2010synthesis}.
The diameter of embedded particles can range from several to several dozen nanometers,
and the characteristic size of microspheres themselves most commonly ranges from tenths to several microns.
One of the most popular applications of magnetic microspheres is 
the magnetic cell separation~--~a technique that allows one to 
magnetically label cells of a specific type and then isolate them
from a heterogeneous cell mixture using a gradient field~\cite{mccloskey2003magnetic,leong2016working}. 
Also microspheres can be used as magnetically controlled carriers for 
targeted drug delivery~\cite{widder1978magnetic,dutz2012microfluidic}
and as force and torque transducers in magnetic tweezers
designed to probe mechanical properties of biomolecules~\cite{amblard1996magnetic,van2015biological}.

Another important class of objects are 
dense three-dimensional (3D) nanoclusters of single-domain particles 
which are sometimes referred to as ``magnetic multicore nanoparticles''~\cite{yang2005effect,dutz2009ferrofluids,schaller2009monte}.
Such clusters are typically covered with a non-magnetic protective coating 
and have a hydrodynamic diameter of 50-200~nm.
Multicore nanoparticles can be thought as intermediate between single-domain nanoparticles
and magnetic microspheres~\cite{dutz2016magnetic}.
From the viewpoint of cell separation, 
nanoclusters have some advantages over micrometer-sized beads:
for example, they are more stable against sedimentation and have a better binding capacity
due to a higher surface-area-to-volume ratio~\cite{leong2016working,xu2011antibody}.
Multicore nanoparticles are considered to be perspective for 
magnetic hyperthermia treatment~\cite{dutz2011magnetic,blanco2015high} 
and magnetic imaging~\cite{lartigue2012cooperative,eberbeck2013multicore}.
Aside from their high biomedical potential, magnetic 3D nanoclusters
are also interesting due to their presence in some types 
of ferrofluids~\cite{buzmakov1996structure,magnet2012haloing}. 
It is known that suspended nanoclusters can significantly alter the fluid's magnetic, mass-transfer and rheological
properties~\cite{ivanov2010magnetophoresis,borin2011ferrofluid}.

For simplicity, it is sometimes assumed that 
microspheres and nanoclusters contain non-interacting and magnetically
isotropic single-domain particles~\cite{amblard1996magnetic,ivanov2010magnetophoresis,borin2011ferrofluid,guo2003preparation}. 
However, in recent years quasi-spherical rigid clusters of different sizes
have been actively studied via numerical simulations~\cite{schaller2009monte,schaller2009effective,usov2016universal,ilg2017equilibrium,weddemann2010dynamic,melenev2010monte,usov2017interaction}. 
And it has been repeatedly demonstrated that interactions between embedded particles as well as their magnetic anisotropy 
can have a noticeable impact on the cluster static~\cite{schaller2009monte,schaller2009effective,usov2016universal,ilg2017equilibrium}
and dynamic~\cite{ilg2017equilibrium,weddemann2010dynamic,melenev2010monte,usov2017interaction} magnetic properties. 
Particularly, in Refs.~\cite{schaller2009monte,ilg2017equilibrium} 
the equilibrium magnetization curve of a quasi-spherical cluster of uniaxial particles was considered. 
Dipole-dipole interactions between particles were taken into account. 
It was demonstrated that for a monodisperse system with a uniform distribution of easy axes
the magnetization is generally lower than predicted by the classical Langevin model
and that both anisotropy and interactions contribute to the decrease of the cluster equilibrium response.

Though a number of simulation results are currently available, 
it can be useful to have an analytical model that links 
properties of particles inside a rigid 3D cluster with the system magnetization.
For single-domain particles dispersed in a liquid matrix, 
many such models exist~\cite{huke2004magnetic}.
Among them the so-called ``modified mean-field theory'' (MMFT) 
remains one of the most widely used due to its simplicity and accuracy~\cite{pshenichnikov1996magneto,ivanov2001magnetic,ivanov2007magnetic}.
In Refs.~\cite{wang2002molecular,pshenichnikov2000equilibrium} it was shown that 
MMFT also gives correct predictions 
for the initial susceptibility of magnetoisotropic particles 
randomly distributed in a solid matrix.
Good agreement between simulations and MMFT 
was obtained for both the bulk system~\cite{wang2002molecular} and 
the finite spherical cluster~\cite{pshenichnikov2000equilibrium}.
The question of whether MMFT is applicable to clusters beyond the weak field limit, 
to the best of our knowledge, has not been addressed in the literature.

In this paper, the equilibrium magnetization curve of a rigid quasi-spherical cluster
of uniaxial particles is studied via Langevin dynamics simulations. 
In contrast to recent works~\cite{schaller2009monte,ilg2017equilibrium},
a special attention is paid to the effect of particle volume fraction on the cluster properties.
The applicability of MMFT for the description of magnetic 3D clusters is tested. 
Possible ways to improve the agreement between the analytical model and simulations are discussed.
 
\section{Model and methods}

\subsection{Model formulation}

Let us consider an ensemble of $N$ identical spherical single-domain particles
randomly distributed within a spherical volume of radius $R$.
Positions of particles inside this volume are fixed, particle overlapping is not allowed.
Each particle has a diameter $d$ and a magnetic moment $\bm{\mu}$,
which can rotate inside the particle,
the corresponding unit vector is $\bm{e} = \bm{\mu}/\mu$.
The magnitude of the magnetic moment is $\mu = M_s v$, 
where $M_s$ is the saturation magnetization of the particle material,
$v = (\pi/6)d^3$ is the particle volume.
Particles have uniaxial magnetic anisotropy, 
which is characterized by the anisotropy constant $K$ 
and the easy axis unit vector $\bm{n}$.
Each particle has its own fixed vector $\bm{n}$.
The orientation distribution of easy axes is random and uniform.
Particles interact with each other via dipole-dipole interactions.
The described system is further referred to as the ``cluster''.
The cluster is immobilized inside a non-magnetic medium and 
subjected to a uniform magnetic field $\bm{H}$
(the corresponding unit vector is $\bm{h} = \bm{H}/H$).
The total magnetic energy of the cluster is
\begin{gather}
	\label{eq:energy} 
		U = U_{Z} + U_{ani} + U_{dd},  \\
	\label{eq:uz}
		U_{Z} = - \mu_0 \mu H \sum_{i = 1}^{N}  \bm{e}_i \cdot \bm{h},    \\
	\label{eq:ua}
		U_{ani} = - K v \sum_{i = 1}^{N}  \left( \bm{e}_i \cdot \bm{n}_i\right)^2, \\
	\label{eq:udd}
		U_{dd} = - \frac{\mu_0 \mu^2}{4 \pi  d^3} 
		\sum_{\substack{i = 1  }}^N \sum_{\substack{j = i + 1}}^N \left[
		\frac{3 (\bm{e}_i \cdot \bm{r}^*_{ij}) (\bm{e}_j \cdot \bm{r}^*_{ij})}{r^{*5}_{ij}} - \frac{\bm{e}_i \cdot \bm{e}_j }{r^{*3}_{ij}}
		\right],
\end{gather}
where $U_{Z}$ is the Zeeman energy, 	
$U_{ani}$ is the magnetic anisotropy energy, 
$U_{dd}$ is the dipole-dipole interaction energy,
the summation in Eqs.~(\ref{eq:uz}--\ref{eq:udd}) is over particles in the cluster,
$\mu_0$ is the magnetic constant,
$\bm{r}^*_{ij} = \bm{r}_{ij}/d$, 
$\bm{r}_{ij}$ is the vector between centers of particles $i$ and $j$.

At non-zero temperature $T$, the normalized magnetic moment of the cluster
\begin{equation}
\bm{m} = \frac{1}{N} \sum_{i = 1}^{N} \bm{e}_i 
\end{equation}
is a random vector with fluctuating magnitude and direction.
The equilibrium magnetization of the cluster can be determined as
\begin{equation}
M = \frac{1}{V}  \left\| \left\langle \sum_{i = 1}^{N} \bm{\mu}_i \right\rangle \right\| = 
M_{\infty} \Big\| \langle \bm{m} \rangle \Big\| = 
 M_{\infty} \langle m_h \rangle,
\end{equation}
where $V = (4\pi/3)R^3$ is the cluster volume, 
$M_{\infty} = \mu N / V$ is the saturation magnetization of the cluster,
$m_h = \bm{m} \cdot \bm{h}$ is the projection of the cluster moment 
on the field direction, angle brackets denote a mean value.
Equilibrium magnetization of the cluster is
determined by several dimensionless parameters.
First of all, this is the so-called Langevin parameter
\begin{equation}
\xi = \frac{\mu_0 \mu H}{k_B T},
\end{equation}
which is the characteristic ratio between Zeeman and thermal energies,
$k_B$ is the Boltzmann constant.
The dependence of $\langle m_h \rangle$ on $\xi$ 
can be considered as the cluster magnetization curve.
Finding this dependency is the main focus of this work.
Other key parameters are
the anisotropy parameter
\begin{equation}
\sigma = \frac{K v}{k_B T},
\end{equation}
the dipolar coupling parameter
\begin{equation}
\lambda = \frac{\mu_0 \mu^2 }{4 \pi d^3 k_B T},
\end{equation}
and the particle volume fraction
\begin{equation}
\varphi = \frac{v N}{V}.
\end{equation}
Let us make some estimates based on material parameters for magnetic solids
given in Ref.~\cite{rosensweig2002heating}.
First of all, it should be noted that dipolar coupling and anisotropy parameters
are not independent variables for particles 
of a given material, $\sigma/\lambda = (24/\mu_0)K/M^2_s$ 
(here we neglect the difference between the particle diameter and the diameter of its magnetic core).
For cobalt ferrite ($M_s = 425~\text{kA~m}^{-1}$, $K = 180$--$200~\text{kJ~m}^{-3}$),
$\sigma/\lambda \simeq 20$;
for magnetite ($M_s = 446~\text{kA~m}^{-1}$, $K = 23$--$41~\text{kJ~m}^{-3}$),
$\sigma/\lambda = 2$--$4$.
Since iron oxide nanoparticles are more common in biomedical applications~\cite{ilg2017equilibrium},
here we confine ourselves to the cases when $\sigma$ and $\lambda$ are comparable.
At $T = 300$~K, magnetite nanoparticles with $d = 10$~nm have $\lambda \simeq 1.3$,
$\xi = 1$ corresponds to $H \simeq 14~\text{kA~m}^{-1}$.
The same nanoparticles with $d = 13$~nm have $\lambda \simeq 2.9$,
$\xi = 1$ corresponds to $H \simeq 6.4~\text{kA~m}^{-1}$.
In this work, the following ranges of control parameters are considered:
$\xi \le 10$, $\sigma \le 10$, $\lambda \le 3$, $\varphi \le 0.3$ and 
$N = 10^2$--$10^3$.

\subsection{Limiting case of non-interacting particles}

Equilibrium magnetic properties of non-interacting uniaxial particles
in a solid matrix were previously discussed in Refs.~\cite{bean1959superparamagnetism,raikher1983magnetization,chantrell1985low,williams1993superparamagnetism,mamiya1998magnetization,cregg1999series}.
Let us briefly recall some results of these works.
If interactions between particles can be neglected, 
i.e. in the limiting cases $\varphi \ll 1$ or $\lambda \ll 1$,
the equilibrium magnetization can be derived within the one-particle approximation.
The ratio between magnetic and thermal energies for an isolated particle
is usually written as
\begin{equation}\label{eq:energy1}
\frac{u}{k_B T} = -\xi \cos \omega + \sigma \sin^2 \vartheta,
\end{equation}
where $\omega$ is the angle between the particle moment 
and the field, $\vartheta$ is the angle between the moment and the easy axis.
The system magnetization is determined by the average value of $\cos \omega = \bm{e} \cdot \bm{h}$,
which can be found as 
\begin{equation} \label{eq:omega_z}
\langle \cos \omega \rangle = \frac{1}{Z}\frac{\partial Z}{\partial \xi},
\end{equation}
where $Z$ is the partition function of the particle.
If particles have negligible magnetic anisotropy ($\sigma \ll 1$),
the partition function is
\begin{equation}
Z = \frac{1}{2} \int_{0}^{\pi} \exp (\xi \cos \omega) \sin \omega d \omega = \frac{\sinh \xi}{\xi},
\end{equation}
which, in combination with Eq.~(\ref{eq:omega_z}), gives the well-known Langevin magnetization:
\begin{equation} \label{eq:lang}
\langle m_h \rangle = \langle \cos \omega \rangle 
=  {\cal L}(\xi) ,
\end{equation}
where ${\cal L}(\xi) \equiv \coth \xi - 1 /\xi $ is the Langevin function.
For uniaxial particles, the partition function and its first derivative 
can be written in the following single-integral forms~\cite{cregg1999series}:
\begin{align}
Z  =  &~{\cal J}_0(\xi,\sigma, \psi)   \nonumber  \\  
\equiv & \int_{0}^{\pi/2}\exp (- \sigma \sin^2 \vartheta) \cosh (\xi \cos \vartheta \cos \psi) \nonumber \\ 
& \times I_0 (\xi \sin \vartheta \sin \psi) \sin \vartheta d \vartheta, \\
\frac{\partial Z}{\partial \xi} = &~{\cal J}_1(\xi,\sigma, \psi)   \nonumber \\ 
\equiv &  \int_{0}^{\pi/2} \exp (- \sigma \sin^2 \vartheta) \Big[ \cosh (\xi \cos \vartheta \cos\psi) \nonumber \\
& \times I_1 (\xi \sin \vartheta \sin \psi) \sin \vartheta \sin \psi + \sinh (\xi \cos \vartheta \cos \psi) \nonumber\\
& \times I_0  (\xi \sin \vartheta \sin \psi) \cos \vartheta \cos \psi\Big] \sin \vartheta d \theta ,
\end{align}
where $\psi$ is the angle between the field and the easy axis ($\cos \psi = \bm{n} \cdot \bm{h}$),
$I_0$ and $I_1$ are the modified Bessel functions of the first kind of order zero and one, correspondingly.
Thus, for an arbitrary particle with a given easy axis orientation $\psi$, 
the following expression is valid:
\begin{equation}\label{eq:omega_uni}
\langle \cos \omega \rangle =  \frac{ {\cal J}_1(\xi,\sigma, \psi)}{{\cal J}_0(\xi,\sigma, \psi)}.
\end{equation}
If particles in the system have different orientations of easy axes,
then one have to average Eq.~(\ref{eq:omega_uni})
over all presented values of $\psi$ to obtain the net magnetization. 
It was demonstrated in Ref.~\cite{raikher1983magnetization} that the distribution of easy axes 
(the system ``orientation texture'') effects the magnetization curve significantly.
For the special case of a random uniform distribution, the magnetization is~\cite{williams1993superparamagnetism,cregg1999series}
\begin{equation}\label{eq:lang_ani}
\langle m_h \rangle = {\cal L}_{ani}(\xi,\sigma) \equiv \int_{0}^{\pi/2} \frac{{\cal J}_1(\xi,\sigma, \psi)}{{\cal J}_0(\xi,\sigma, \psi)} \sin \psi d \psi . 
\end{equation}
The integral Eq.~(\ref{eq:lang_ani}) is denoted here as ${\cal L}_{ani}(\xi,\sigma)$.
This function can be considered as a generalization of the standard Langevin function
for the case of solid dispersions with random orientation texture.
In the limit of negligible anisotropy, two functions coincide, i.e. ${\cal L}_{ani}(\xi, 0) = {\cal L}(\xi)$.
For finite non-zero values of $\xi$~and~$\sigma$, 
${\cal L}_{ani}(\xi, \sigma) < {\cal L}(\xi)$~\cite{raikher1983magnetization}.
However, the zero-field slope of the magnetization curve (the initial magnetic susceptibility~$\chi$)
does not depend on $\sigma$ ~\cite{bean1959superparamagnetism,chantrell1985low}:
\begin{multline}
\chi = \frac{M}{H}\bigg|_{H \rightarrow 0} 
= \frac{\mu_0 \mu^2 \varphi}{ v k_B T } \Bigg(\frac{{\cal L}_{ani}(\xi,\sigma)}{\xi}\bigg|_{\xi \rightarrow 0}\Bigg)   \\
= \frac{\mu_0 \mu^2 \varphi}{ v k_B T }\Bigg(\frac{{\cal L}(\xi)}{\xi}\bigg|_{\xi \rightarrow 0}\Bigg) 
= 3 \chi_L \Bigg(\frac{{\cal L}(\xi)}{\xi}\bigg|_{\xi \rightarrow 0}\Bigg)
= \chi_L 
\end{multline} 
where $\chi_L = \mu_0 \mu^2 \varphi / 3 v k_B T = 8 \lambda \varphi$ 
is the so-called Langevin susceptibility.
For infinite anisotropy $\sigma = \infty$ and finite values of $\xi$,
magnetic moments of particles can be considered as Ising-like spins
with only two available states $\vartheta = 0$ and $\vartheta~=~\pi$~\cite{mamiya1998magnetization}.
The magnetization in this asymptotic limit is given~by~\cite{cregg1999series}
\begin{equation}\label{eq:lang_spin_int}
{\cal L}_{ani}(\xi,\infty) = \int_{0}^{\pi/2} \cos \psi \tanh (\xi \cos \psi) \sin \psi d \psi. 
\end{equation}

\subsection{Dipole-dipole interactions and modified mean-field theory}

When one considers a body homogeneously filled with particles interacting 
via long-range dipole-dipole interactions, 
one of the main things that should be taken into account is the demagnetizing field. 
If a magnetizable body is placed in a uniform magnetic field $\bm{H}$,
then the field inside the body $\bm{H}_{int}$ does not coincide
with $\bm{H}$ in the general case.
The difference between $\bm{H}$ and $\bm{H}_{int}$ is known as the demagnetizing field,
it is created by the surface divergence of the body's own magnetization $\bm{M}$~\cite{joseph1965demagnetizing}.
For an arbitrary shaped body, demagnetizing fields can have a complex spatial distribution.
But for the special case of an
ellipsoid, the demagnetizing field is uniform.
If $\bm{H}$ lies along one of the principal axes of a magnetizable ellipsoid,
then $\bm{H}_{int}$ and $\bm{M}$ also lie along this direction.
Magnitudes of these vectors are connected as
\begin{equation}\label{eq:hint}
H_{int} = H - \kappa M,
\end{equation}  
where  $0 \le \kappa \le 1$ is the demagnetizing factor of the ellipsoid along the chosen axis.
The factor $\kappa$ depends only on the shape of the ellipsoid and not on its size.
For an infinitely elongated (needle-like) ellipsoid parallel to the field, $\kappa = 0$,
and for a sphere it is $\kappa = 1/3$.
 
Now let us consider a needle-like body with $H~=~H_{int}$ ($\kappa~=~0$),
filled with interacting magnetoisotropic particles ($\sigma = 0$). 
Even in this case, the equilibrium magnetic response
can not be described by the Langevin model.
A possible way to expand the model is the well-known Weiss mean-field theory. 
According to it, an effective magnetic field acting locally on an arbitrary particle
consists of the applied field $H$ and 
an additional term which describes the impact of the particle surroundings.
This term is proportional to the system magnetization $M$,
the proportionality factor is normally equal to the Lorentz value~$1/3$~\cite{kittel1996introduction}.
The system magnetization is then given by
\begin{equation}\label{eq:weiss}
M = M_L (H + M/3),
\end{equation}
where $M_L(H) \equiv M_{\infty}{\cal L}(\mu_0 \mu H / k_B T)$.
However, Eq.~(\ref{eq:weiss}) is known to overestimate the effect 
of dipole-dipole interactions on concentrated assemblies of single-domain particles.
Particularly, the Weiss theory predicts a spontaneous transition 
into an orientationally ordered ``ferromagnetic'' state at $\chi_L = 3$~\cite{cebers1982thermodynamic,zhang1995spontaneous},
but such transition has not been observed experimentally.
Some more advanced theories and numerical simulations 
indicate the possibility of the transition 
both for liquid~\cite{wei1992orientational} and solid~\cite{klapp2001ferroelectric} matrices,
but corresponding critical values of $\chi_L$ are significantly larger than predicted by the Weiss theory.
In Ref.~\cite{pshenichnikov1996magneto} the following heuristic modification of the mean-field
theory was proposed for dispersions of single-domain
particles in a liquid matrix (i.e., for ferrofluids):
\begin{equation}\label{eq:mmft1}
M = M_L \left(H + M_L(H)/3\right).
\end{equation}
In this expression, the impact of the system on an arbitrary particle
is described not by the system actual magnetization $M$,
but by the magnetization the system would have in the absence of interactions, i.e. by~$M_L(H)$. 
The statistical-mechanical approach developed in Ref.~\cite{ivanov2001magnetic}
subsequently justified the validity of the heuristic formula Eq.~(\ref{eq:mmft1}).
Moreover, the authors of Ref.~\cite{ivanov2001magnetic} suggested its refined version that reads
\begin{equation}\label{eq:mmft2}
M = M_L\left(H + \frac{M_L(H)}{3} + \frac{M_L(H)}{144}\frac{dM_L(H)}{dH}\right).
\end{equation}
Eqs.~(\ref{eq:mmft1}) and (\ref{eq:mmft2}) are now known as the first- and second-order
modified mean-field theories, correspondingly (MMFT1 and MMFT2).
At small and moderate values of $\lambda$ and $\varphi$,
they are both in good agreement with experimental and numerical results 
on ferrofluid magnetization, though MMFT2
has a wider range of applicability ~\cite{ivanov2007magnetic}.
However, MMFTs assume a homogeneous distribution of particles
in the system and hence fail to describe an enhanced magnetic response
at strong coupling $\lambda \ge 3$, which is due to the formation
of chain-like aggregates~\cite{ivanov2004applying}.
The applicability of MMFTs for solid magnetic dispersions,
where the formation of aggregates is forbidden,
was numerically investigated in Refs.~\cite{pshenichnikov2000equilibrium,wang2002molecular}.
Only the initial magnetic susceptibility of the solid system was considered.
According to Ref.~\cite{wang2002molecular}, 
MMFT1 describes $\chi$ well for $\lambda \le 3$ and $\varphi \le 0.25$,
while MMFT2 slightly overestimates the susceptibility.
The applicability of MMFT for solid systems at non-zero fields is to be tested.
Using previously defined dimensionless parameters,
Eqs.~(\ref{eq:mmft1}) and (\ref{eq:mmft2}) can be rewritten in the form
\begin{equation}\label{eq:mmft}
\langle m_h \rangle = {\cal L}\Big(\xi + {\cal C}_{mf}(\xi){\cal L}(\xi)\Big),
\end{equation}
where $ {\cal C}_{mf}$ is the mean-field parameter,
which can depend on the applied field in the general case. 
For MMFT1:
\begin{equation}\label{eq:cmf1}
 {\cal C}_{mf} = \chi_L,
\end{equation}
for MMFT2:
\begin{equation}\label{eq:cmf2}
{\cal C}_{mf}(\xi) = \chi_L \left(1 + \frac{\chi_L}{16}\frac{d {\cal L}(\xi)}{d \xi}\right).
\end{equation}
For a body with $\kappa \neq 0$, $H$ in magnetization expressions must be replaced by $H_{int}$. 
For a sphere, the magnetization curve can be then obtained in the following parametric form:
\begin{gather}
\langle m_h \rangle = {\cal L}\Big(\xi_{int} + {\cal C}_{mf}(\xi_{int}){\cal L}(\xi_{int})\Big), \label{eq:mmft_cl}\\
\xi = \xi_{int} + \chi_L \langle m_h \rangle,\label{eq:xi_int}
\end{gather}
where $\xi_{int} = \mu_0 \mu H_{int}/k_B T$ is the parameter ($0~\le~\xi_{int}~<~\infty$), Eq.~(\ref{eq:xi_int}) 
corresponds to Eq.~(\ref{eq:hint}) with $\kappa = 1/3$.

To describe the cluster of interacting uniaxial particles,
we propose here the following phenomenological generalization of Eq.~(\ref{eq:mmft_cl}),
where both Langevin functions are replaced by ${\cal L}_{ani}$:
\begin{equation}\label{eq:mmfta}
\langle m_h \rangle = {\cal L}_{ani}\Big(\xi_{int} + {\cal C}_{mf}(\xi_{int}){\cal L}_{ani}(\xi_{int},\sigma),\sigma\Big),
\end{equation}
The replacement of the first Langevin function ensures the correct behavior 
in the limit of non-interacting particles ($\xi_{int} = \xi$, ${\cal C}_{mf} = 0$). 
As for the second replacement, we here speculate that
the impact of a randomly textured solid dispersion 
on an arbitrary particle can be described by the mean-field term proportional to ${\cal L}_{ani}$,
just like the impact of a system of magnetoisotropic particles is described by ${\cal L}$ in MMFT.
A suitable choice of the function ${\cal C}_{mf} = {\cal C}_{mf}(\xi)$ 
in Eq.~(\ref{eq:mmfta}) is discussed in Sec.~\ref{sec:ani_cluster}.

\subsection{Langevin dynamics simulation}

To check the accuracy of the described models, the Langevin dynamics simulation is used.
The Langevin equation that describes the magnetodynamics of a single-domain particle 
is the stochastic Landau-Lifshitz-Gilbert equation~\cite{ilg2017equilibrium,garcia1998langevin}.
For the $i$th particle of the simulated cluster it reads
\begin{equation}\label{eq:llg1}
	\frac{d \bm{\mu}_i}{d t} = - \gamma \left[\bm{\mu}_i \times \bm{H}^{tot}_i\right] - \frac{\gamma \alpha}{\mu} \left[\bm{\mu}_i \times \left[\bm{\mu}_i \times \bm{H}^{tot}_i \right]\right],
\end{equation}
where $\gamma = \gamma_0 / (1 + \alpha ^2)$, $\gamma_0$ is the gyromagnetic ratio (in meters per ampere per second), 
$\alpha$ is the dimensionless damping constant, 
$\bm{H}^{tot}_i = \bm{H}^{det}_i + \bm{H}^{fl}_i$,
$\bm{H}^{det}_i = - \left(\partial U/\partial \bm{\mu}_i\right)/\mu_0$ is the total deterministic field acting on the particle,
$\bm{H}^{fl}_i$ is the fluctuating thermal field.
$\bm{H}^{fl}_i(t)$ is a Gaussian stochastic process 
with the following statistical properties:
\begin{equation}
\left\langle H^{fl}_{i,k}(t)\right\rangle =  0, \:
\left\langle H^{fl}_{i,k}(t_1) H^{fl}_{j,l}(t_2) \right\rangle =  2 D \delta_{ij}\delta_{kl}\delta(t_1 - t_2), 
\end{equation}
where $k$ and $l$ are Cartesian indices,
$D = \alpha k_B T/\mu_0 \mu\gamma (1+\alpha^2)$.
Eq.~(\ref{eq:llg1}) can be rewritten in the dimensionless form:
\begin{gather}\label{eq:llg}
\frac{d \bm{e}_i}{d t^*} = - \frac{1}{2 \alpha}\left[\bm{e}_i \times \bm{\xi}^{tot}_i\right] - \frac{1}{2} \left[\bm{e}_i \times \left[\bm{e}_i \times \bm{\xi}^{tot}_i\right]\right],
\end{gather}
where the $t^* = t/\tau_D$ is the dimensionless time,
$\tau_D = \mu_0 \mu / 2 \alpha \gamma k_B T$ is the characteristic time scale
of the rotary diffusion of the magnetic moment,
$\bm{\xi}^{tot}_i = \mu_0 \mu \bm{H}^{tot}_i/k_B T = \bm{\xi}^{det}_i + \bm{\xi}^{fl}_i$,
\begin{equation}
\bm{\xi}^{det}_i  = \xi \bm{h} + 2 \sigma (\bm{e}_i \cdot \bm{n}_i) \bm{n}_i 
+ \lambda \sum_{j \neq i}^{N} \left[ \frac{3  \bm{r}^*_{ij} (\bm{e}_j \cdot \bm{r}^*_{ij})}{r^{*5}_{ij}} - \frac{\bm{e}_j }{r^{*3}_{ij}} \right],
\end{equation}
\begin{equation}
\left\langle \xi^{fl}_{i,k}(t^*)\right\rangle =  0, \:
\left\langle \xi^{fl}_{i,k}(t^*_1)\xi^{fl}_{j,l}(t^*_2) \right\rangle =  \frac{4 \alpha^2}{1 + \alpha^2}\delta_{ij}\delta_{kl}\delta(t^*_1 - t^*_2). 
\end{equation}

\begin{figure}
	\includegraphics{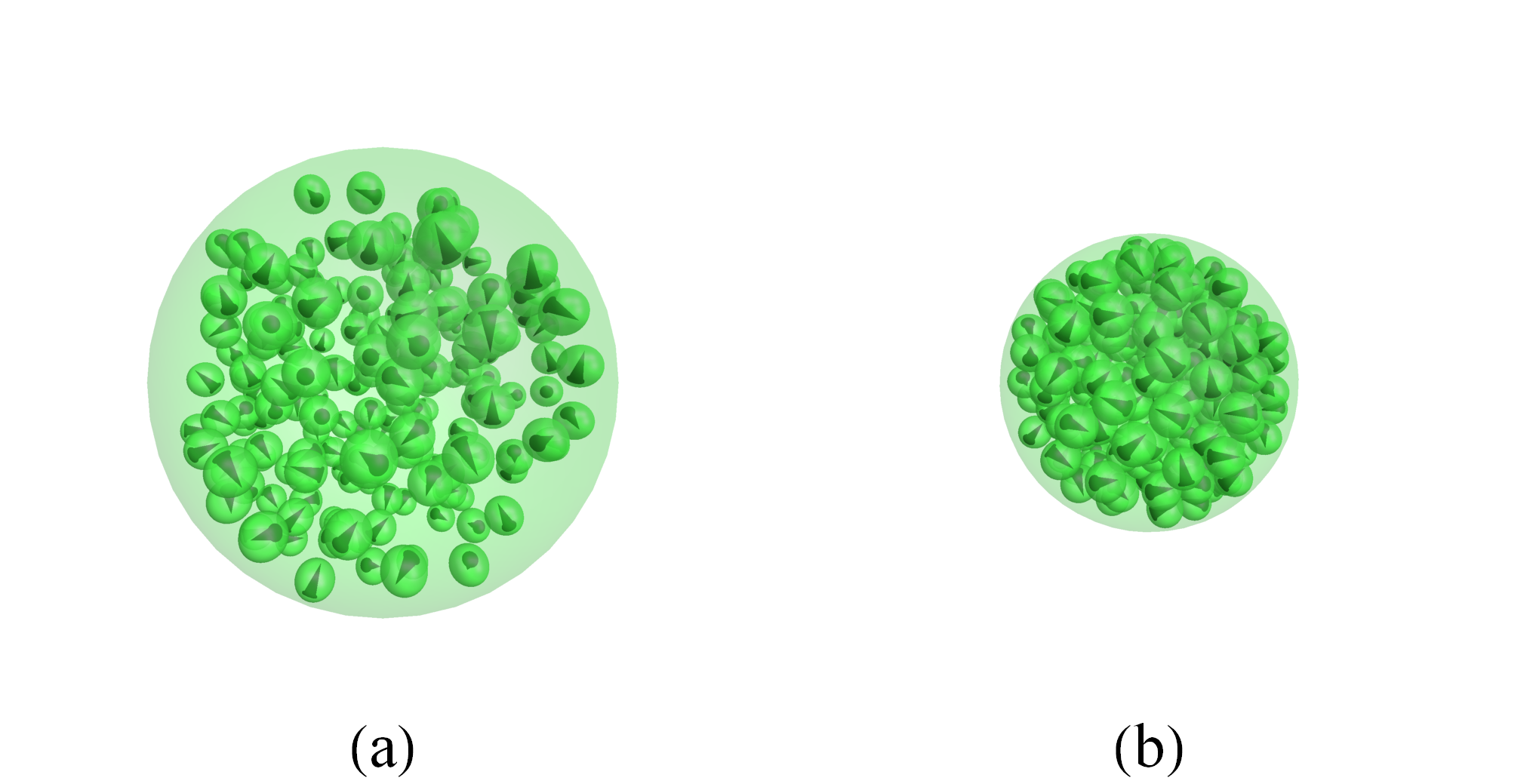}
	\caption{\label{fig:cluster_snpashots}
		Examples of rigid clusters used in simulations, $\lambda = \sigma = \xi = 0$, $N = 200$.
		(a) $\varphi = 0.1$, $R \approx 6.3d$,
		(b) $\varphi = 0.3$, $R \approx 4.4d$.}
\end{figure}

The input parameters of the simulation are $N$, $\varphi$, $\lambda$, $\xi$ and $\sigma$. 
The cluster at given $N$ and $\varphi$ is generated as follows.
The $i$th particle is randomly placed inside a cube with a side length of $2R$
($1 \le i \le N$, $R = (d/2)\sqrt[3]{N/\varphi}$).
If after this the particle is outside of the sphere of radius $R$ or if 
it overlaps with previously placed particles (i.e., with particles $j < i$), 
the position is rejected and the new position is generated. 
This is repeated until a suitable position is found.
Then $\bm{n}_i$ and the initial state of $\bm{e}_i$ are chosen at random. 
Then the state of the particle $i + 1$ is generated according to the same rules.
Examples of clusters with $N = 200$ and different volume fractions are shown in Fig.~\ref{fig:cluster_snpashots}.

After the cluster is generated, the Heun scheme~\cite{garcia1998langevin} is used for the numerical integration of Eq.~(\ref{eq:llg})
the damping constant is $\alpha = 0.2$,
the integration time step is $\Delta t^* = 0.002$, unless otherwise specified.
Dipole-dipole interaction fields between particles in the cluster are calculated without truncation, 
no periodic boundary conditions (PBCs) are applied.
The main result of the simulation is the average normalized magnetization of the cluster $\langle m_h \rangle$.
In the case $\sigma = 0$, the sampling of $m_h$ values typically starts after the time $t^* = 100$,
but for $\sigma > 0$ a much longer equilibration period might be required.
This issue is discussed in Sec.~\ref{sec:ani_cluster}.
For each particular set of input parameters, the magnetization value is additionally	
averaged over several independent realizations of the cluster.
However, the results for different realizations are proved to be close,
so their number is not large.
Most of the magnetization curves presented below are averaged over 
ten realizations of the cluster.

In addition to clusters, this paper also briefly discusses the equilibrium 
magnetization of a bulk solid dispersion of magnetic nanoparticles (see Sec.~\ref{sec:bulk}).
The input simulation parameters in this case are the same as for the cluster.
The simulation cell is a cube with a side length of $L = d\sqrt[3]{\pi N / 6 \varphi}$. 
PBCs are applied in all three directions.
The dipolar fields in the system are calculated using the standard Ewald summation 
with ``metallic'' boundary conditions. 
This technique ensures a proper handling of long-range effects of dipole-dipole 
interactions. In its ``metallic'' version the internal field in the simulation box
coincides exactly with the applied field.
A detailed description of the technique is available in Refs.~\cite{wang2002molecular,wang2001estimate}.

\section{Results and discussion}

\subsection{Magnetically isotropic particles in a bulk solid matrix}\label{sec:bulk}

\begin{figure*}
	\includegraphics{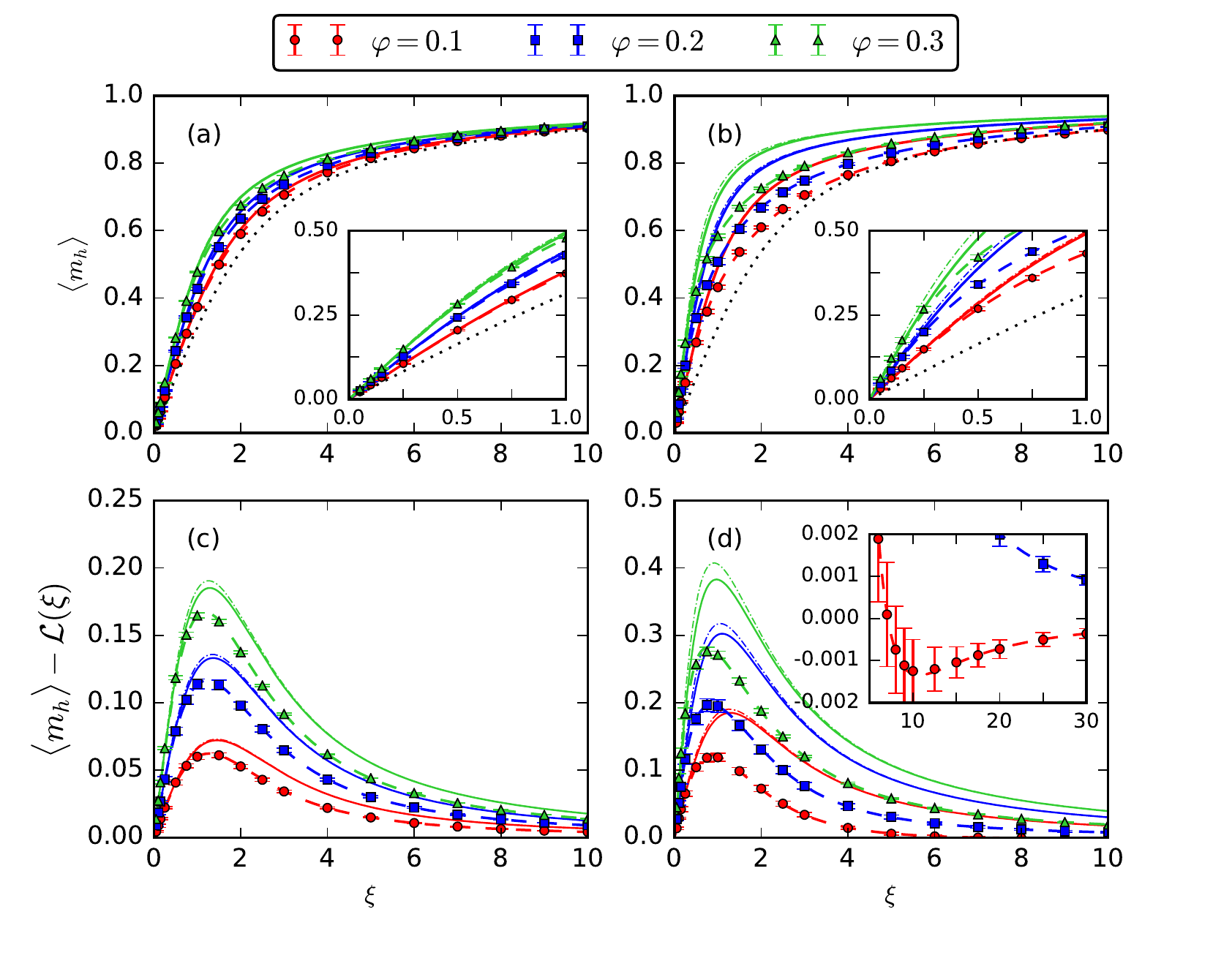}
	\caption{\label{fig:mc_bulk}
		Equilibrium magnetization curves of bulk solid systems with $H = H_{int}$ and $\sigma = 0$.
		$\lambda = 1$ (a) and $3$ (b). 
		Symbols are the simulation results for $N = 1000$, 
		different combinations of symbols and colors 
		correspond to different particle volume fractions $\varphi$ (see legend). 
		Solid curves are MMFT1 results 
		for the same values of $\varphi$ [Eqs.~(\ref{eq:mmft}) and (\ref{eq:cmf1})],
		dot-dashed lines are MMFT2 results [Eqs.~(\ref{eq:mmft}) and (\ref{eq:cmf2})],
		dashed curves are from the analytical model given by Eqs.~(\ref{eq:mmft}) and~(\ref{eq:cpade}),
		dotted lines are from the Langevin model [Eq.~(\ref{eq:lang})].
		Insets in (a) and (b) zoom on the weak field region.
		Figures (c) and (d) show differences between the 
		magnetization values shown in (a), (b) and the Langevin function; 
		(c) corresponds to (a) (i.e., to $\lambda = 1$) and (d) corresponds to (b) (i.e., to $\lambda = 3$).
		The inset in (d) shows values of $\langle m_h\rangle - {\cal L}(\xi)$  
		in the strong field region.}
\end{figure*}
Before moving on to the main object of our interest, 
i.e. the finite-size magnetic cluster,
it may be useful to consider the equilibrium magnetization 
of a bulk solid matrix filled with magnetic nanoparticles
and to test the applicability of MMFTs for such system.
In numerical simulations we model bulk in a standard way by applying PBCs to a cubic simulation cell. 
First of all, the usage of PBCs minimizes possible size effects that may arise 
in the simulation of the cluster. 
Besides, we use the ``metallic'' version of the Ewald summation technique to calculate dipole-dipole interactions.
This method assumes that the large system formed by the simulation cell and its PBC-images 
is surrounded by a medium with infinite magnetic permeability~\cite{wang2002molecular}.
In this case, $\xi = \xi_{int}$ and the system magnetic behavior is the same as that of an elongated cylindrical sample.
So, the demagnetizing fields, which are inevitable for the quasi-spherical cluster in a non-magnetic medium, are now absent.
In this section, we only consider the case $\sigma = 0$.

\begin{figure}
	\includegraphics[scale=1]{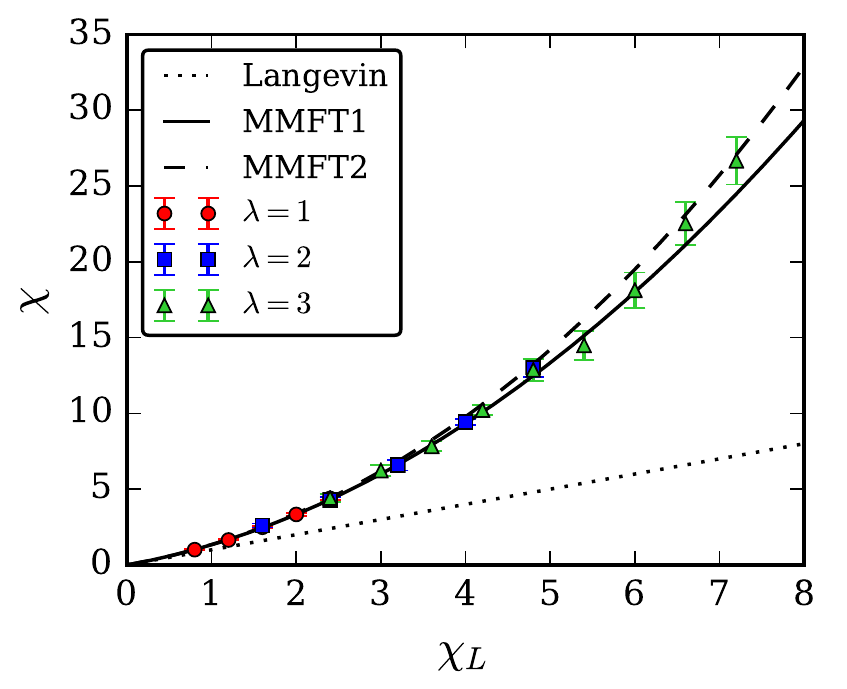}
	\caption{\label{fig:chi_bulk}
		Initial magnetic susceptibility $\chi$ of magnetoisotropic particles 
		embedded in a bulk solid matrix vs. the Langevin susceptibility $\chi_L$.
		Symbols are the simulation results for $N = 1000$,
		different symbols correspond to different values of 
		the dipolar coupling constant $\lambda$ (see legend).
		Dotted line corresponds to the Langevin model ($\chi = \chi_L$),
		solid line is from MMFT1 [Eq.~(\ref{eq:chi:mmft1})],
		dashed line is from MMFT2 [Eq.~(\ref{eq:chi:mmft2})].
		}
\end{figure}

Static magnetization curves of bulk systems with different values of $\varphi$ and $\lambda$
are given in Fig.~\ref{fig:mc_bulk}.
To emphasize the effect of interparticle interactions on the equilibrium magnetic properties,
we also give in Figs.~\ref{fig:mc_bulk}(c)~and~\ref{fig:mc_bulk}(d) the differences between the system actual
magnetization and the Langevin function.
Symbols denote values obtained after averaging over ten independent 
realizations of the simulated system, 
error bars here and below denote corresponding 95\% confidence intervals.
It is seen that MMFT1 [Eqs.~(\ref{eq:mmft}) and (\ref{eq:cmf1})]
and MMFT2 [Eqs.~(\ref{eq:mmft}) and (\ref{eq:cmf2})]
give good agreement with the simulation data in the weak field range ($\xi < 1$).
In Fig.~\ref{fig:chi_bulk} the system initial susceptibility $\chi$ 
is plotted vs. the Langevin susceptibility $\chi_L$.
According to MMFT1, the susceptibility is~\cite{ivanov2001magnetic}
\begin{equation}\label{eq:chi:mmft1}
\chi = \chi_L (1 + \chi_L/3),
\end{equation}
and according to MMFT2, it is
\begin{equation}\label{eq:chi:mmft2}
\chi = \chi_L (1 + \chi_L/3 + \chi_L^2/144).
\end{equation}
The susceptibility of the simulated system is estimated simply as 
$\chi = 3 \chi_L \langle m_h (\xi = 0.05) \rangle/0.05$.
For $\lambda \le 3$ and $\varphi \le 0.25$, 
MMFT1 describes calculated susceptibilities well, which 
agrees with the results of Ref.~\cite{wang2002molecular}.
At $\lambda = 3$ and $\varphi > 0.25$ 
(which corresponds to $\chi_L > 6$ for $\lambda = 3$), 
the susceptibility is seemingly better described by MMFT2.
In Ref.~\cite{wang2002molecular} the behavior of solid systems 
at $\varphi > 0.25$ was not investigated.
More conspicuous deviations between MMFT1 and the simulation results 
are observed in Fig.~\ref{fig:mc_bulk} at moderate and strong fields. 
The theory clearly overestimates the simulation results 
at $\xi \ge 1$ for all inspected values of interaction parameters.
The deviation is larger for higher~$\lambda$.
As seen in Figs.~\ref{fig:mc_bulk}(c)~and~\ref{fig:mc_bulk}(d), 
the magnetization of a bulk solid system at moderate and large $\xi$
is closer to the Langevin curve than MMFT1 predicts.
Despite this fact, the deviation between the simulation results and 
the Langevin model is still significant. 
The maximum difference between $\langle m_h \rangle$
and ${\cal L}(\xi)$ is observed at $\xi \sim 1$.
For $\lambda = 3$ and $\varphi = 0.3$, it reaches $\approx 0.28$.
In other words, the difference between the non-reduced magnetization $M$
and $M_L$ is $\approx 28\%$ of the system saturation magnetization~$M_{\infty}$.
As $\xi$ increases, the calculated magnetization approaches the Langevin 
curve much faster than it should according to MMFT1.
For one of the investigated parameter sets ($\lambda = 3$, $\varphi = 0.1$),
the calculated values of $\langle m_h\rangle$ are even smaller than ${\cal L}(\xi)$ 
at $\xi \gtrsim 7$ (though the maximum value of the difference $M_L - M$ 
is less than one percent of $M_{\infty}$ as seen in the inset of Fig.~\ref{fig:mc_bulk}(d)).
As for MMFT2, it overestimates simulation data at large fields even stronger than MMFT1.
Such overestimation was not observed in ferrofluid simulations --
in the strong coupling case ($\lambda > 2$) and at $\xi \ge 1$ 
the ferrofluid magnetization is either slightly lower than predictions of MMFTs 
(at high concentrations) or greatly exceeds it (at low concentrations)~\cite{wang2002molecular}.
A possible explanation is as follows.
According to MMFTs, the effective field $H_{eff}$ acting on an arbitrary particle $i$ 
is always larger than the applied field $H$ and the difference $H_{eff} - H$ 
becomes larger with increasing $H$.
Within this theory, dipole-dipole interactions between the $i$th particle
and its surroundings, on  average, always help the particle to align with the applied field.
Based on our simulation results, 
this is true for solid dispersions of magnetic particles in the weak field limit.
But at large fields the situation can become complicated 
due to the anisotropic nature of dipole-dipole interactions.
Let us choose a Cartesian coordinate system 
so that its center coincides with the center of the $i$th particle
and the $Z$ axis coincides with the applied field direction $\bm{h}$.
If $\xi$ is large enough, magnetic moments of all particles in the system 
are predominantly directed along the $Z$ axis. 
If the particle $j$ with $\bm{e}_j || \bm{h}$
is placed somewhere on the $Z$ axis, then the dipolar field created by this particle 
at the location of the $i$th particle is co-directed with $\bm{h}$.
However, if the $j$th particle is placed in the $XY$ plane, then its 
dipolar field at the $i$th particle location is directed opposite to $\bm{h}$.
In ferrofluids, the anisotropy of dipole-dipole interactions
results in the field-induced anisotropy of the pair distribution function\cite{elfimova2012theory}. 
In a liquid matrix, the probability to find the $j$th particle on the $Z$ axis 
in the close contact with the particle $i$ becomes higher 
with increasing $\xi$. 
Two co-directed particles with $\bm{r}_{ij} || \bm{h}$ tend
to attract each other and form an energetically favorable 
``head-to-tail'' configuration.
This effect is noticeable even at relatively low 
dipolar coupling $\lambda \simeq 1$.
At large $\lambda$, it transforms in the well-known 
formation of chain-like aggregates.
On the contrary, the probability to find the $j$th
particle in the plane $XY$ in the close contact
with the particle $i$ decreases with increasing $\xi$.
Two co-directed particles with $\bm{r}_{ij} \perp \bm{h}$ 
tend to repel each other.
So, as the field increases, magnetic particles in a liquid 
matrix tend to redistribute themselves so that 
the local surroundings of the $i$th particle is more likely 
to contain particles that favor the orientation of $\bm{e}_i$ along $\bm{h}$.
But in our case, the isotropic spatial distribution of particles is frozen.
So, at large applied fields the $i$th particle
is surrounded both by particles that help it to align with the field
and by particles that interfere with such behavior. 
It seems probable that, as the average result of such competition,
the effective field $H_{eff}$ acting on the $i$th particle in a solid
matrix with increasing $H$ becomes smaller than the corresponding effective field
in a liquid matrix.
As the inset in Fig.~\ref{fig:mc_bulk}(d) suggests,
in some cases $H_{eff}$ can even become slightly smaller than $H$.
In Eq.~(\ref{eq:mmft}) the effect of the particle
surroundings is controlled by the mean-field parameter ${\cal C}_{mf}$.
In order to correctly describe the observed behavior 
of a solid dispersion, this parameter should become significantly lower
than the standard MMFT1 value $\chi_L$ at large fields.
\begin{figure}
	\includegraphics[scale=1]{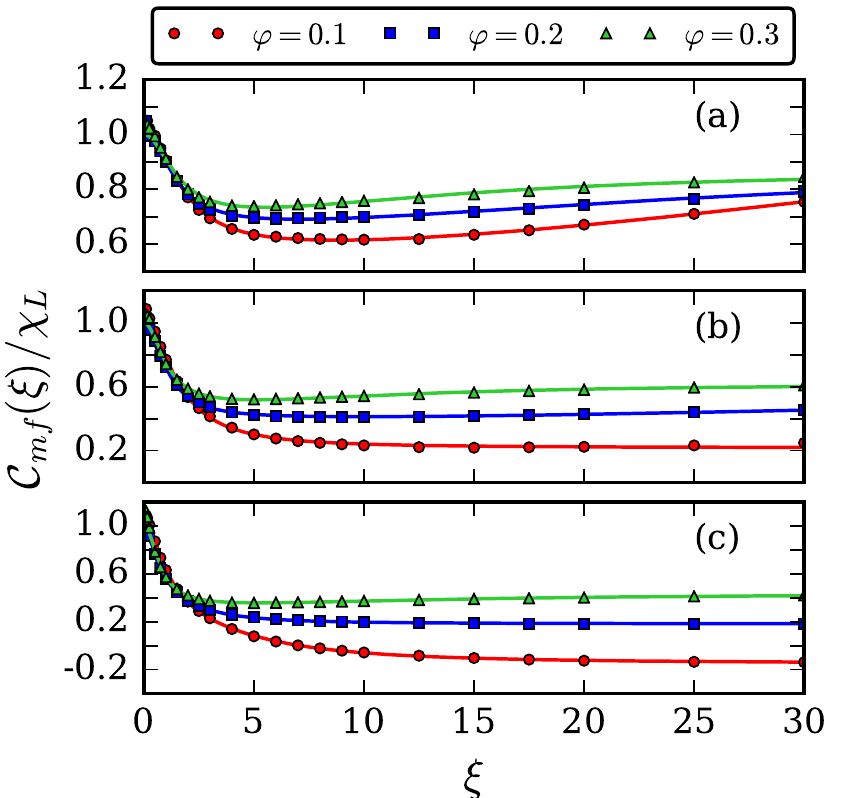}
	\caption{\label{fig:mfcoef}
		Applied field dependencies of the mean-field parameter 
		for bulk solid systems of magnetoisotropic particles.
		$\lambda = 1$ (a), 2 (b) and 3 (c).
		Symbols are simulation results for $N = 1000$, 
		different symbols correspond to different volume fractions (see legend).
		Solid lines are from the approximation~Eq.~(\ref{eq:cpade}).
		}
\end{figure}
Figure~\ref{fig:mfcoef} shows the values of ${\cal C}_{mf}$ 
extracted from the simulation data using the expression
\begin{equation}\label{eq:c_extraction}
{\cal C}_{mf}(\xi) = \frac{{\cal L}^{-1}\big( \langle m_h \rangle \big) - \xi}{{\cal L}(\xi)},
\end{equation}
where ${\cal L}^{-1}(x)$ is the inverse Langevin function (its values where obtained numerically).
The mean-field parameter in the figure is dived by $\chi_L$.
It is seen that at $\xi \ll 1$ ${\cal C}_{mf}/\chi_L \gtrsim 1$,
but it becomes lower at large fields, just as expected.
At $\xi \lesssim 2$, values of ${\cal C}_{mf}/\chi_L$ decrease relatively fast,
at a given $\lambda$ they are almost the same for different volume fractions.
For $\xi > 2$, the quantity ${\cal C}_{mf}/\chi_L$ seemingly begins to reach a plateau.
The values of ${\cal C}_{mf}/\chi_L$ at large fields for different combinations
of $\lambda$ and $\varphi$ do not coincide.
Particularly, they are very different for $\lambda = 1$ and $\varphi = 0.3$
and for $\lambda = 3$ and $\varphi = 0.1$, despite the fact that 
the Langevin susceptibility is the same in both cases ($\chi_L = 2.4$).
For $\lambda = 3$ and $\varphi = 0.1$, the mean-field parameter at large $\xi$
becomes negative, which is why $\langle m_h \rangle$ at these parameters becomes
smaller than ${\cal L}(\xi)$.
At large $\xi$, values of ${\cal C}_{mf}/\chi_L$ increase with increasing $\varphi$
if $\lambda$ is fixed. The increase in $\lambda$ at a fixed volume fraction
has the opposite effect -- in this case ${\cal C}_{mf}/\chi_L$ decreases.
To be able to check whether or not the mean-field parameters obtained
for bulk systems are applicable for the description of clusters at large fields,
we approximated the dependencies presented in Fig.~\ref{fig:mfcoef}
with the expression
\begin{equation}\label{eq:cpade}
{\cal C}_{mf}(\xi) = \chi_L \frac{1 + a_2 \xi^2 + a_4 \xi^4}{1 + b_{2} \xi^2 + b_{4} \xi^4 }.
\end{equation}
Eq.~(\ref{eq:cpade}) contains only even powers of $\xi$,
so that Eq.~(\ref{eq:mmft}) remains an odd function of the magnetic field.
Coefficients $a_2$, $a_4$, $b_2$ and $b_4$ were separately determined 
for each investigated combination of $\lambda$ and $\varphi$
using non-linear least squares fitting. The calculated values 
are given in Table~\ref{tab}. The approximations are valid at least
up to $\xi = 30$.
\begin{table}
	\caption{\label{tab} 
		Fitting parameters of Eq.~(\ref{eq:cpade}) for different 
		dipolar coupling constants and particle volume fractions.}
	\begin{tabular}{c c | c c c c}
		\hline
		$\lambda$ & $\varphi$ & $a_2$ & $a_4$ & $b_2$ & $b_4$ \\
		\hline \hline
		1 & 0.1 & 0.165553 & 0.000119 & 0.286021 & 0.000083 \\
		  & 0.2 & 0.294838 & 0.000583 & 0.444823 & 0.000662 \\
		  & 0.3 & 0.289164 & 0.001879 & 0.417613 & 0.002166 \\
		\hline
		2 & 0.1 & 0.312283 & 0.015637 & 0.629365 & 0.071964 \\
		  & 0.2 & 0.340067 & 0.000143 & 0.854058 & 0.000202 \\
		  & 0.3 & 0.429252 & 0.004477 & 0.901514 & 0.007247 \\
		\hline
		3 & 0.1 & 0.217007 & -0.004704 & 0.864708 & 0.031570 \\
		  & 0.2 & 1.632664 & 0.128207 & 3.249253 & 0.708223 \\
		  & 0.3 & 0.555831 & 0.004696 & 1.689059 & 0.010885 \\
		\hline \hline
	\end{tabular}
\end{table}

\subsection{Cluster of magnetically isotropic particles}\label{sec:iso_cluster}

\begin{figure}[b]
	\includegraphics[scale=1]{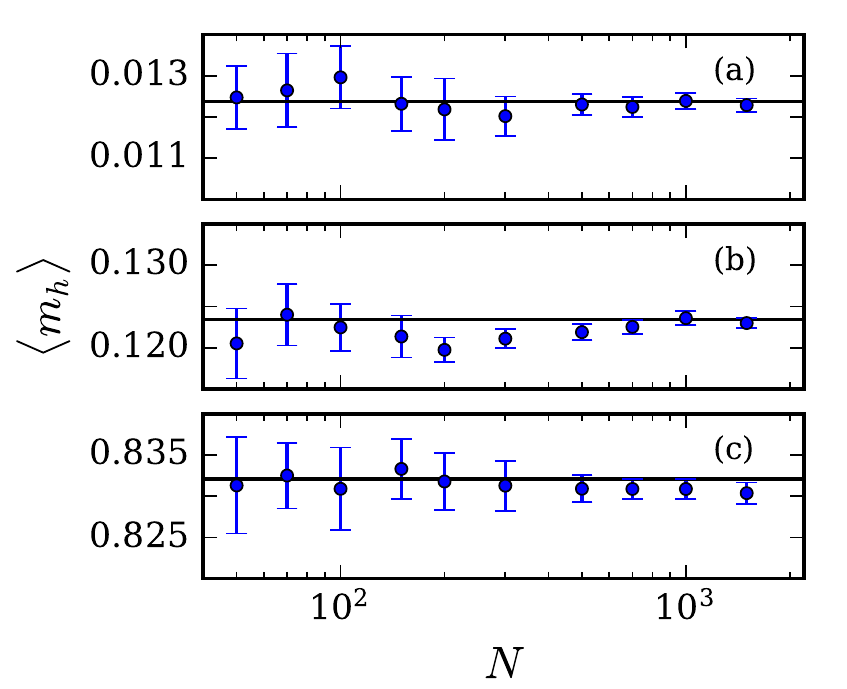}
	\caption{\label{fig:size_effects}
		Normalized equilibrium magnetization of the cluster vs. 
		the number of particles it contains. 
		$\lambda = 3$, $\varphi = 0.3$ and $\sigma = 0$.
		$\xi = 0.1$ (a), 1 (b), 10 (c).
		Symbols are simulation results.
		Horizontal lines correspond to predictions of
		the analytical model given by Eqs.~(\ref{eq:mmft_cl})~and~(\ref{eq:cpade}).
	}
\end{figure}
One of our main concerns regarding simulations of the cluster 
were possible finite-size effects. 
In Ref.~\cite{wang2003boundary} it was shown that  
properties of finite spherical containers with ferrofluid 
depend heavily on the system size in the case of strong dipolar coupling.
Equilibrium magnetization of systems with $N \sim 10^2$--$10^3$ 
proved to be much smaller than corresponding thermodynamic limit values. 
Magnetization values of rigid clusters with $\sigma = 0$ 
are shown in Fig.~\ref{fig:size_effects} as a function of
the particle number $N$ at different values of $\xi$.
These data are calculated for $\lambda = 3$ and
$\varphi = 0.3$, i.e. for the largest values of interaction parameters
considered in this work. 
Luckily, the results do not indicate strong 
size dependencies for rigid quasi-spherical clusters.
This gives hope that approximations Eq.~(\ref{eq:cpade}) 
derived for a bulk system will work for small clusters as well.

\begin{figure}	
	\includegraphics{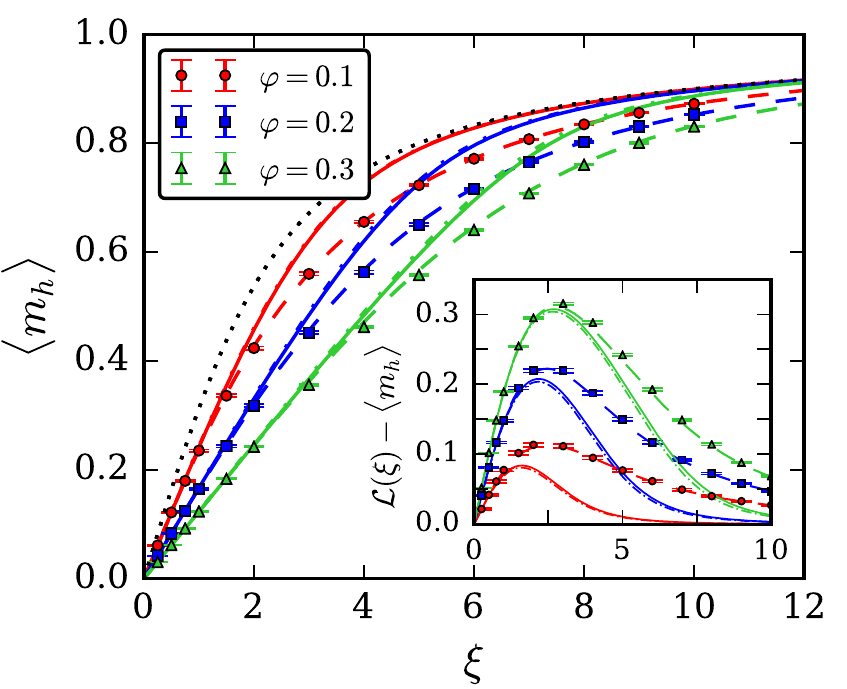}
	\caption{\label{fig:mc_iso}
		Equilibrium magnetization curves of clusters with $\sigma = 0$ and $\lambda = 3$.
		Symbols are simulation results for $N = 1000$, 
		different combinations of symbols and colors correspond to different volume fractions (see legend).
		Solid cures are prediction of MMFT1 [Eqs.~(\ref{eq:cmf1}) and (\ref{eq:mmft_cl})] 
		for the same volume fractions,
		dot-dashed curves are predictions of MMFT2 [Eqs.~(\ref{eq:cmf2}) and (\ref{eq:mmft_cl})].
		Dashed lines are from the analytical model given by Eqs.~(\ref{eq:mmft_cl})~and~(\ref{eq:cpade}).
		Dotted line is the Langevin function [Eq.~(\ref{eq:lang})].
		The inset shows corresponding differences between the Langevin function
		and magnetization values.}
\end{figure}

Static magnetization curves for clusters with $\sigma = 0$ and $\lambda = 3$
are given in Fig.~\ref{fig:mc_iso}.
Due to the presence of demagnetizing fields, 
the effect of interactions is the opposite of what was observed in the previous section.
Magnetization of the cluster is now always smaller than the Langevin model predicts
and the cluster equilibrium response is weaker the higher the volume fraction $\varphi$.
Just like in the bulk case, MMFT1 
(which is now given by Eqs.~(\ref{eq:cmf1}) and (\ref{eq:mmft_cl}))
provides an accurate description of the magnetization curve initial slope, 
but overestimates the simulation results at strong fields ($\xi \gtrsim 2$). 
MMFT2 [Eqs.~(\ref{eq:cmf2}) and (\ref{eq:mmft_cl})] again gives 
larger magnetization values than MMFT1,
but it should be noted that the difference between two theories 
is much less pronounced than in the bulk case.
The combination of Eq.~(\ref{eq:mmft_cl}) and approximation Eq.~(\ref{eq:cpade}) 
with fitting parameters taken from Table~\ref{tab} 
accurately describes the cluster magnetization
at all investigated values of $\varphi$ and $\xi$.
The foregoing is also true for smaller coupling parameters,
but the difference between the cluster magnetization
and the Langevin function in this case is much less distinguishable.
For example, at $\lambda = 1$ and $\varphi = 0.3$, the maximum
value of ${\cal L} (\xi) - \langle m_h \rangle$ is $\approx 0.09$.
For $\lambda = 3$ and $\varphi = 0.3$, the difference
${\cal L} (\xi) - \langle m_h \rangle$
can become larger than~$0.3$
(which is seen in the inset of Fig.~\ref{fig:mc_iso}).

\subsection{Cluster of uniaxial particles}  \label{sec:ani_cluster}

\begin{figure}[b]
	\includegraphics{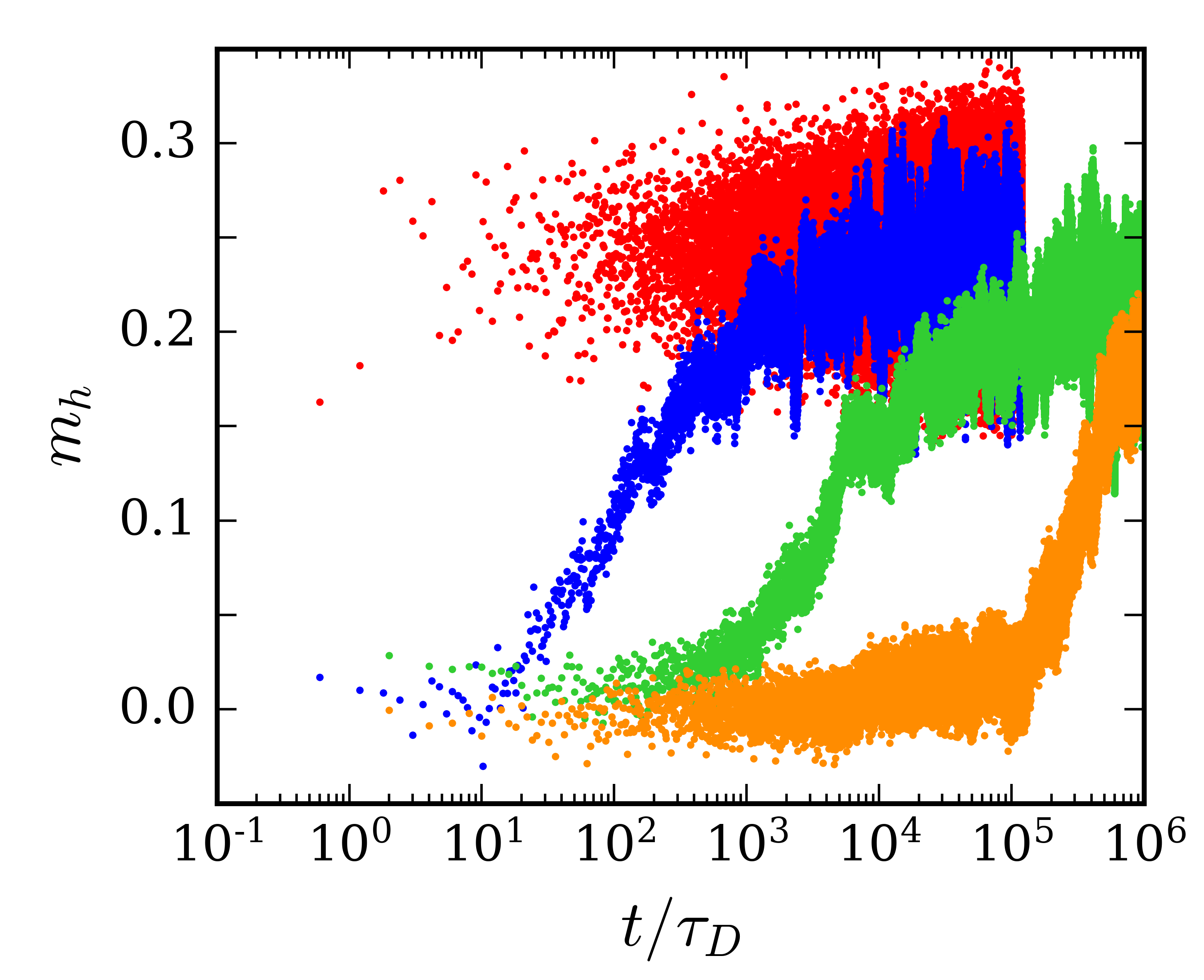}
	\caption{\label{fig:magdyn}
	Instantaneous values of the cluster normalized magnetization vs. simulation time.
	Simulation results for $N = 400$, $\lambda = 1$, $\varphi = 0.1$ and $\xi = 1$.
	From top to bottom: $\sigma = 0$, 10, 15 and 20.
	}
\end{figure}
\begin{figure}
	\includegraphics{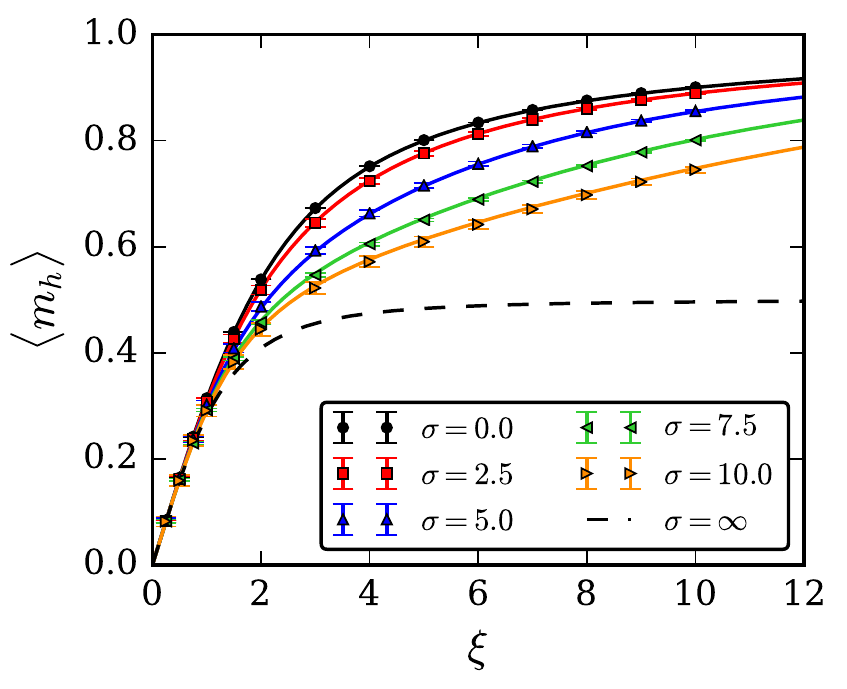}
	\caption{\label{fig:mcani_nonint}
	Equilibrium magnetization curves of clusters of non-interacting ($\lambda = 0$)
	uniaxial particles. Symbols are simulation results for $N = 400$,
	different combinations of symbols and colors 
	correspond to different anisotropy parameters $\sigma$ (see legend).
	Lines are predictions of Eq.~(\ref{eq:lang_ani}) for the same 
	values of $\sigma$.
	}		
\end{figure}

One can expect that the time necessary 
for the cluster to reach the equilibrium magnetization value
from the initial random state
will increase with increasing anisotropy parameter $\sigma$.
The reason is that magnetic moments of particles 
will have to overcome the anisotropy energy barrier.
In zero magnetic field and in the absence of interactions,
the characteristic time scale that determines 
how long it will take for the magnetic moment to overcome the barrier 
(i.e., to spontaneously change its orientation from $\bm{n}$
to energetically equivalent $-\bm{n}$) 
is called the N{\'e}el relaxation time ($\tau_N$).
This time increases exponentially with increasing $\sigma$.
With a good accuracy $\tau_N$ is given by the approximation~\cite{coffey1994simple}
\begin{equation} \label{eq:neel}
\tau_N = \tau_D \frac{e^{\sigma} - 1}{2 \sigma} \left[\frac{1}{1 + 1/\sigma}\sqrt{\frac{\sigma}{\pi}} + 2^{-\sigma - 1}\right]^{-1}.
\end{equation}
In the limit of negligible anisotropy ($\sigma \ll 1$), 
the N{\'e}el time $\tau_N$ reduces to the relaxation time $\tau_D$.
Eq.~(\ref{eq:neel}) gives 
$\tau_N \simeq 14 \tau_D$ for $\sigma = 5$, 
$\tau_N \simeq 6.8 \times 10^2 \tau_D$ for $\sigma = 10$,
$\tau_N \simeq 5.3 \times 10^4 \tau_D$ for $\sigma = 15$ and
$\tau_N \simeq 5.0 \times 10^6 \tau_D$ for $\sigma = 20$.
Figure~\ref{fig:magdyn} demonstrates a very similar non-linear slow down.
This figure shows the dynamics of the cluster magnetization 
for different values of $\sigma$ at a fixed 
field $\xi = 1$. 
In the beginning $m_h$ is close to zero,
but then it starts to increase and gradually reaches a non-zero equilibrium value.
It is seen that as $\sigma$ varies from 0 to 20,
the characteristic equilibration period increases by several orders of magnitude.
For larger fields ($\xi > 1$), the period decreases, but the direct simulation 
of clusters with $\sigma > 10$ still remains challenging from a computational viewpoint. 
Due to the restrictions of available computational resources,
only cluster with $N = 400$ and $\sigma \le 10$ are considered below.
The integration time step is slightly increased to $\Delta t^* = 0.003$.
Further increase of the time step can potentially lead to erroneous simulation results~\cite{ilg2017equilibrium}.
We use equilibration period $t^* = 500$ for $2.5 \le \sigma \le 7.5$
and $t^* = 4000$ for $\sigma = 10$.

Magnetization curves of a cluster of non-interacting  
uniaxial particles ($\lambda = 0$) were first calculated as a test.
The results are given in Fig.~\ref{fig:mcani_nonint}.
Calculations are in full agreement with Eq.~(\ref{eq:lang_ani}).
The linear response at weak fields is always the same as for the Langevin 
model, but for $\xi > 1$ the growth of $\langle m_h \rangle$ is slower
the higher the anisotropy parameter $\sigma$.

\begin{figure*}
	\includegraphics{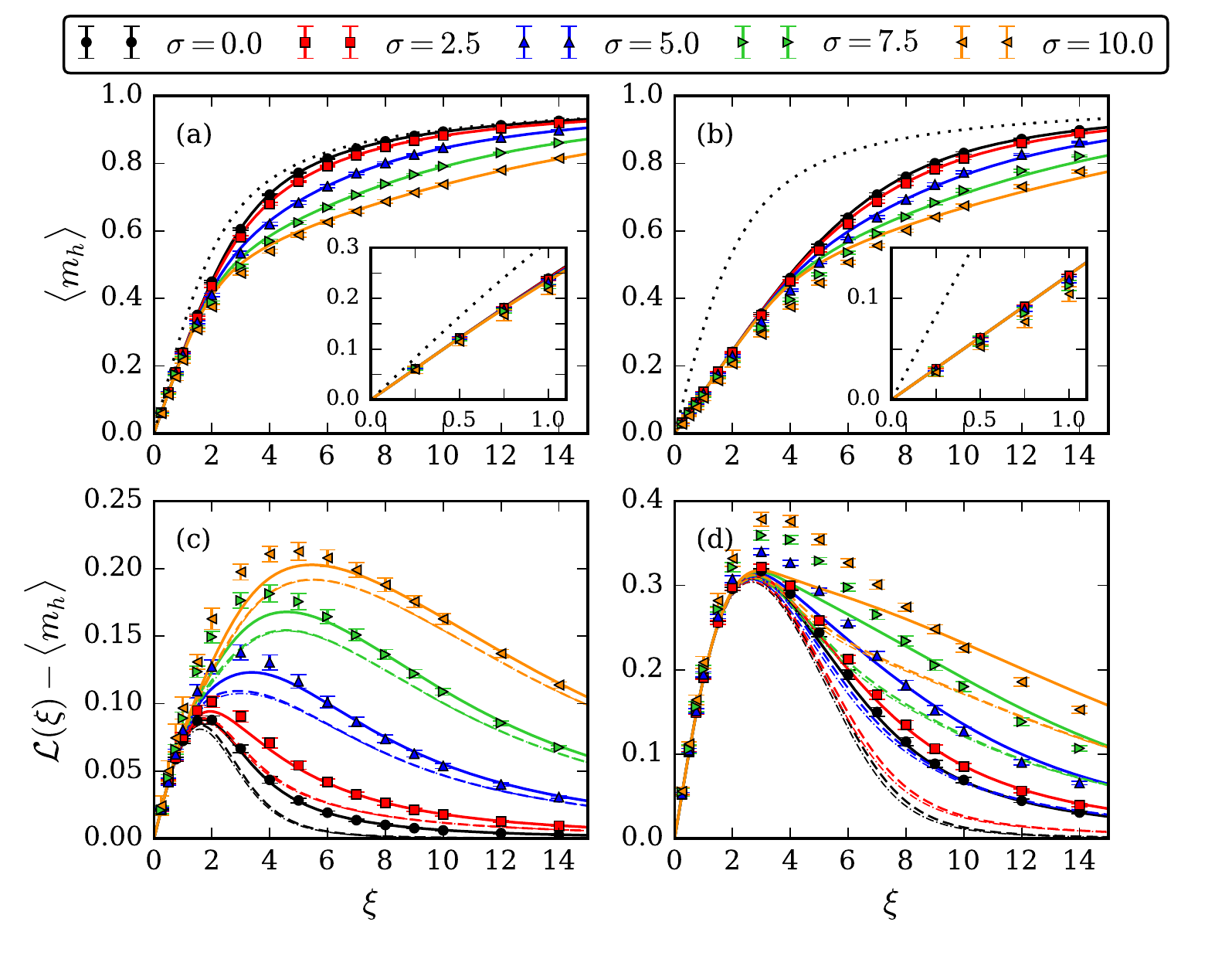}
	\caption{\label{fig:mc_ani}
		Equilibrium magnetization curves of clusters with $\varphi = 0.3$.
		$\lambda = 1$ (a) and $3$ (b). 
		Symbols are the simulation results for $N = 400$, 
		different combinations of symbols and colors correspond to different anisotropy parameters $\sigma$ (see legend). 
		Solid curves are predictions of the analytical model given by Eqs.~(\ref{eq:mmfta}) and (\ref{eq:cpade})
		for the same values of $\sigma$.
		Dotted lines are from the Langevin model Eq.~(\ref{eq:lang}).
		Insets in (a) and (b) zoom on the weak field region.
		Figures (c) and (d) show differences between the Langevin function 
		and the magnetization values shown in (a)~and~(b);
		(c) corresponds to (a) (i.e., to $\lambda = 1$) and (d) corresponds to (b) (i.e., to $\lambda = 3$).
		Dashed lines in (c) and (d) correspond to the analytical model given by Eqs.~(\ref{eq:cmf1}) and~(\ref{eq:mmfta}),
		dot-dashed lines correspond to the analytical model given by Eqs.~(\ref{eq:mmfta}) and~(\ref{eq:cmf2_ani}).}
\end{figure*}
Magnetization curves for clusters with $\sigma \le 10$, $\varphi = 0.3$ and different 
values of $\lambda$ are given in Fig.~\ref{fig:mc_ani}.
The most noticeable effect of increasing anisotropy,
just as in the case of non-interacting particles,
is the saturation slowdown at large fields.
Our phenomenological modification of MMFT given by Eq.~(\ref{eq:mmfta})
correctly reproduces this feature. 
As a mean-field parameter in Eq.~(\ref{eq:mmfta}), we use 
approximation Eq.~(\ref{eq:cpade}) with fitting
parameters previously obtained for bulk magnetoisotropic systems
(see Table~\ref{tab}).
For $\lambda = 1$ (Figs.~\ref{fig:mc_ani}(a) and \ref{fig:mc_ani}(c)),
the combination of Eqs.~(\ref{eq:mmfta}) and (\ref{eq:cpade}) 
shows great quantitative agreement with simulation data.
The largest deviations are observed at intermediate fields $2 \lesssim \xi \lesssim 6$,
where the analytical model overestimates the magnetization of simulated clusters.
For $\sigma = 10$, the largest deviation is $\approx 0.03$. 
Deviations become more pronounced at $\lambda = 3$ 
(Figs.~\ref{fig:mc_ani}(b) and \ref{fig:mc_ani}(d)).
At $\sigma = 10$, the largest deviation is now $\approx 0.07$. 
Besides, at $\lambda = 3$ and $\xi \gtrsim 10$, calculated magnetizations
become larger than predictions of the analytical model.
Figs.~\ref{fig:mc_ani}(c) and \ref{fig:mc_ani}(d)
additionally demonstrate the predictions of
``anisotropic generalizations'' of MMFT1 and MMFT2.
For MMFT1, this is simply a combination of Eqs.~(\ref{eq:cmf1})
and~(\ref{eq:mmfta}).
For MMFT2, the mean-field parameter Eq.~(\ref{eq:cmf2}) 
was modified using the same intuitive approach, which was
used to obtain Eq.~(\ref{eq:mmfta}) -- the function 
${\cal L}(\xi)$ was replaced with ${\cal L}_{ani}(\xi, \sigma)$:
\begin{equation}\label{eq:cmf2_ani}
{\cal C}_{mf}(\xi, \sigma) = \chi_L \left(1 + \frac{\chi_L}{16}\frac{\partial {\cal L}_{ani}(\xi,\sigma)}{\partial \xi}\right).
\end{equation}
It is seen that ``generalized'' MMFTs overestimate calculated magnetizations at all field values
starting from $\xi \simeq 2$.
Just like in Sec.~\ref{sec:iso_cluster}, 
the predictions of MMFT2 only slightly exceed the predictions of MMFT1.

\section{Conclusions}

In this work, equilibrium magnetization curves of a 
random quasi-spherical cluster of single-domain nanoparticles are studied 
numerically and analytically.
Langevin dynamics simulations show that, due to dipole-dipole interactions
between particles, magnetization of the cluster is 
generally lower than predicted by the classical Langevin model.
This is in full agreement with recent findings of Refs.~\cite{schaller2009monte,ilg2017equilibrium}.
It is shown that, in the case of negligible magnetic anisotropy and 
weak applied fields, magnetization curves can be successfully described by
the so-called modified mean-field theory, initially proposed for the description
of concentrated ferrofluids. 
However, as the field increases, the theory starts to overestimate 
the cluster magnetization. 
The discrepancy can be minimized by adjusting the mean-field parameter of MMFT,
so that it decreases with increasing applied field.
The explicit form of the dependency between the mean-field parameter ${\cal C}_{mf}$
and the Langevin parameter $\xi$ (which determines the impact of the applied field on the system) 
is turned out to be different for different 
values of the dipolar coupling parameter $\lambda$ and the particle volume fraction $\varphi$.
For some specific combinations of $\lambda$ and $\varphi$,
dependencies ${\cal C}_{mf} = {\cal C}_{mf}(\xi)$ are obtained
in the form of approximation formulas.
Clearly, finding a universal dependency ${\cal C}_{mf} = {\cal C}_{mf}(\xi, \lambda, \varphi)$
would be useful from a practical point of view, 
but this requires a rigorous statistical mechanical treatment of the problem,
which is beyond the scope of this work.

It is also shown that if particles have non-negligible anisotropy
(characterized by the anisotropy parameter $\sigma$)
and the distribution of their easy axes is random and uniform, then, 
at given values of $\xi$, $\lambda$ and $\varphi$, the magnetization
of the cluster decreases with increasing $\sigma$.
The decrease is much stronger at large fields.
For weak dipolar coupling ($\lambda \simeq 1$),
this effect can be accurately taken into account simply by replacing
all Langevin functions ${\cal L}(\xi)$ in the magnetization expression given by MMFT [Eq.~(\ref{eq:mmft_cl})] 
with its generalization ${\cal L}_{ani}(\xi, \sigma)$ [Eq.~(\ref{eq:lang_ani})].
Function ${\cal L}_{ani}(\xi, \sigma)$ is the exact solution for magnetization of
non-interacting uniaxial particles with random orientation texture.
At larger coupling parameters ($\lambda > 1$), such simple approach
demonstrates noticeable quantitative deviations from the simulation results.

In this work, only  monodisperse systems are considered.
But it is known that magnetization of rigid clusters 
can also be influenced by the polydispersity of particles~\cite{schaller2009monte}.
The combined effect of magnetic anisotropy, interparticle interactions
and polydispersity on static magnetization curves of finite-size quasi-spherical clusters
will be considered in future papers.    

\section{Acknowledgments}

The work was supported by Russian Science Foundation (project No.~17-72-10033).
The author is grateful to Prof.~A.~F.~Pshenichnikov for valuable discussions.

\bibliography{lib}

\end{document}